\documentclass{vgtc}                          

\ifpdf
  \pdfoutput=1\relax                   
  \usepackage{graphicx}                
  \DeclareGraphicsExtensions{.pdf,.png,.jpg,.jpeg} 
\else
  \usepackage{graphicx}                
  \DeclareGraphicsExtensions{.eps}     
\fi%

\graphicspath{{figures/}{pictures/}{images/}{./}} 

\usepackage{microtype}                 
\PassOptionsToPackage{warn}{textcomp}  
\usepackage{textcomp}                  
\usepackage{mathptmx}                  
\usepackage{times}                     
\usepackage{cite}                      
\usepackage{tabu}                      
\usepackage{booktabs}                  
\usepackage{gensymb}
\usepackage{subfig}
\usepackage{hyperref}

\usepackage[usenames, dvipsnames]{color}

\onlineid{0}

\vgtccategory{Research}

\vgtcinsertpkg

\title{Visual Analysis of Spatio-Temporal Event Predictions: \\Investigating the Spread Dynamics of Invasive Species}

\author{
	Daniel Seebacher\thanks{e-mail: daniel.seebacher@uni.kn}\\ %
		\scriptsize University of Konstanz %
	\and Johannes H\"au{\ss}ler\thanks{e-mail: johannes.3.haeussler@uni.kn}\\ %
		\scriptsize University of Konstanz %
	\and Michael Hundt \thanks{e-mail: michael.hundt@uni.kn}\\ %
		\scriptsize University of Konstanz
	\and Manuel Stein\thanks{e-mail: manuel.stein@uni.kn}\\ %
		\scriptsize University of Konstanz\\
    \and \mbox{}
    \and \mbox{}
    \and Hannes M{\"u}ller\thanks{e-mail: hannes.mueller@lubw.bwl.de}\\ %
		\scriptsize LUBW
	\and Ulrich Engelke\thanks{e-mail: ulrich.engelke@data61.csiro.au}\\ %
		\scriptsize CSIRO Data61
	\and Daniel Keim\thanks{e-mail: daniel.keim@uni.kn}\\ %
		\scriptsize University of Konstanz
}%

\abstract{Invasive species are a major cause of ecological damage and commercial losses. A current problem spreading in North America and Europe is the vinegar fly Drosophila suzukii. Unlike other Drosophila, it infests non-rotting and healthy fruits and is therefore of concern to fruit growers, such as vintners.  Consequently, large amounts of data about infestations have been collected in recent years. However, there is a lack of interactive methods to investigate this data. We employ ensemble-based classification to predict areas susceptible to infestation by D. suzukii and bring them into a spatio-temporal context using maps and glyph-based visualizations. Following the information-seeking mantra, we provide a visual analysis system \textit{Drosophigator} for spatio-temporal event prediction, enabling the investigation of the spread dynamics of invasive species. We demonstrate the usefulness of this approach in two use cases.
}

\CCScatlist{ 
  \CCScat{I.3.3}{Computer Graphics}%
{Picture/Image Generation}{Display algorithms};
  \CCScat{H.5.2}{Information Interfaces
and Presentation}{User Interfaces}{User-centered design}
}

\begin{document}



\maketitle

\section{Introduction} 

Non-native plants, fungus or animal species that out-compete native species often cause severe economic and ecological damage to our planet. With increasing globalization through trade and travel routes, humankind has created opportunities for invasive species to establish themselves in new regions all over the earth. 

An exemplary invasive insect currently spreading around Europe and North America is the Asian vinegar fly \textit{Drosophila suzukii} or spotted wing Drosophila (D. suzukii). In 2008, first occurrences were reported in California, Spain, and Italy rapidly followed by other regions and countries~\cite{cini2012review,cini2014tracking}. In contrast to other Drosophila species, D. suzukii infests even non-rotting and healthy fruits. It has a wide range of possible host plants that have thin-skinned fruits, like cherries, berries or grapes. An adult female fly can lay 1-10 eggs per fruit and 200-400 eggs within its lifespan of 8-25 days. Depending on temperature and other external factors, these eggs become adult flies within 11-24 days. Thus, 13-15 generation cycles are possible during one year. As a result of the spread of D. suzukii, the USA, for example, noted an annual loss of \$500 million~\cite{asplen2015invasion} in fruit production within a few years. \textit{Agroscop}, the Swiss center of excellence for agricultural research, has also published data of crop losses from 2014~\cite{kaiser2015drosphila} showing that in some Swiss cantons, 80-100\% of cherries were unmarketable. Consequently, industry and science are tirelessly searching for novel ways to keep the spread of D. suzukii under control through a better understanding of their spread behavior. Institutes such as the European and Mediterranean Plant Protection Organization~\footnote{EPPO - \url{https://gd.eppo.int}} or the State Viticulture Institute (WBI) in Freiburg run global databases with weekly to monthly reports about present threats and new findings. Focused on the data gathering aspect these systems are, however, often analytically limited to providing simple D. suzukii distribution maps.  
To this end, various approaches have been proposed to explore the recorded data. Wiman et al.~\cite{wiman2014integrating}, e.g., make use of the fact that insects are ectotherms, which means that their body temperature equals the ambient temperature. Therefore, low temperatures are a key cause of insect overwinter mortality. The authors tried to estimate D. suzukii populations in different life stages, based on average daily temperatures of some specific fruit production sites combined with trap catches and fruit infestation counts. With their temperature model they found some confirmation of population trends with trap data, and to a limited extent with fruit infestation data. 
Building on top of this work, other proposed approaches try to optimize temporal and spatial dislocation of control measures by conducting studies on D. suzukii's plasticity of cold tolerance and its overwinter behavior~\cite{jakobs2015adult,rossi2016multiple}. Spatial and temporal dislocation is caused by mainly measuring in high ripening seasons and at orcharding sites. Focusing on temperature alone neglects the environmental aspects under which the fly could best procreate, or survive even in colder seasons. Other approaches focus on several integrated pest management (IPM) strategies instead. An extensive review of current methods as well as a categorization is given by Haye et al.~\cite{haye2016current}. They introduce strategies that focus on chemical, cultural~\cite{cormier2015exclusion, del2017cost} or biological control~\cite{daane2016first,wang2016foraging}.


The multitude of approaches shows that analyzing the spread of invasive species is a complex problem. There are many different external influences, which affect the spread of D. suzukii, such as surrounding areas, time, temperature, food supply and many more. This is aggravated by the fact, that these influences have to be considered in a temporal and geospatial context. This illustrates the need of researchers for interrogating large amounts of complex empirical evidence interactively, in order to gain insights. 

In this paper we present our application \textit{Drosophigator} (\textit{Drosophila Investigator}). We follow a visual analytics approach for interactive exploration of large amounts of heterogeneous data sources, including trap counts of D. suzukii, surrounding high-detail land use data, and related metadata. To help researchers investigate the spread dynamics of invasive species, we proceed as follows: First we train an ensemble of classifiers to predict time and place of possible infestations by D. suzukii. These infestation events, are cumulatively visualized with a glyph-based visualization and brought into a spatio-temporal context by placing them on a map. By allowing zoom and filter capabilities, as well as details on demand our application enables domain experts to understand the spread dynamics of invasive species. We demonstrate the usefulness of \textit{Drosophigator} in two use cases.

\section{Data Description} \label{sec:background}

We performed several expert interviews with the State Viticulture Institute (WBI) in Freiburg, Germany, in order to gain a better understanding of the influences and factors about the spread of D. suzukii as well as to identify current challenges faced by domain experts. The WBI offers through their web service VitiMeteo\footnote{VitiMeteo - \url{http://www.vitimeteo.de/}} forecast models for different fungi species, monitoring data for various pests, as well as weather data related to viticulture in the federal state of Baden-W\"{u}rttemberg. In our interviews we found that although a lot of data about D. suzukii is being collected by the WBI they lack adequate methods to analyze and interpret the arising amounts of data as well as visualization techniques to communicate and present related findings. 

\begin{figure}[!ht]
	\centering
	\includegraphics[width=\columnwidth]{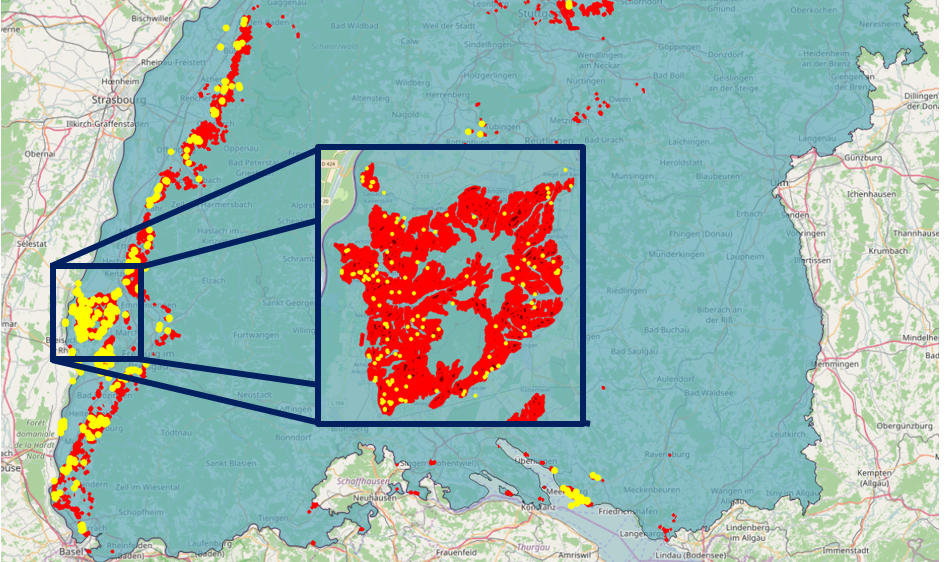}
	\caption{Vineyards (highlighted in red) in Baden-W\"{u}rttemberg, as well as measurements stations by the WBI (highlighted in yellow). Highlighted is the Kaiserstuhl, one of the biggest wine regions in Baden-W\"{u}rttemberg.}
	\label{img:vineyards_bawu}
\end{figure}

In the data provided by VitiMeteo are, among other things, observations of the spread of D. suzukii. This data consists of trap findings of D. suzukii as well as percentage information about how many berries were infested in a sample taken at the station. Additionally, there is percentage information about how many eggs were found in a sample. This percentage can be over 100~\%, if there are more egg findings than berries in a sample. These observations are collected from 867 stations non-uniformly spread over Baden-W\"{u}rttemberg as shown in Figure~\ref{img:vineyards_bawu}. Some of them only report observations for one day, others report multiple observations over a time period of up to 1641 days. The observations are rather sparse and irregularly sampled, which makes the use of standard time series analysis techniques challenging, if not impossible. Consequently, Drosophigator should enable researchers of the WBI to interactively analyze this complex data source. 

The Julius-K{\"u}hn-Institute~\cite{jki} (JKI) suggests that the number of trap findings are increasing in late summer and stay high until winter. Additionally, Pelton et al.~\cite{Pelton2016} found that areas surrounded by woodland exhibit an earlier infestation. As a result, the focus of our application is the analysis of the spread dynamics, exemplified by D. suzukii, by taking temporal distribution as well as environmental factors into account. In order to test the hypotheses of the JKI and Pelton et al., we gathered the relevant data from different resources. The time of year is already present in our observation data provided by the WBI. To gather the height of every measuring station, we make use of the ASTER Global Digital Elevation Map~\cite{aster} which was released by the Ministry of Economy, Trade, and Industry (METI) of Japan and the United States National Aeronautics and Space Administration (NASA). For the land coverage, the State Institute for Environment, Measurements and Nature Conservation of the Federal State of Baden-W\"{u}rttemberg (LUBW), provided us with data from the ATKIS~\cite{grunreich1992atkis} project for the state of Baden-W\"{u}rrtemberg. This includes high-detail statewide land usage information. It consists of main groups, such as forests or industry, but also subgroups, such as coniferous forest or treatment plans. Overall there are 83 different combinations of groups and subgroups.

\section{Ensemble-based classification of infested areas}\label{sec:classification}


To identify regions, in our case vineyards in Baden-W\"{u}rttemberg, which are potentially endangered by D. suzukii, we use machine learning to train a model using the data provided by the WBI in combination with the data collected from ATKIS and ASTER.  This allows us to learn which combination of features make areas, at certain points in time, susceptible to infestation. By applying the trained model on other areas we can find new potentially endangered areas.

\subsection{Data Preparation}
When training our model we need to determine which areas are severely affected by D. suzukii and which are not. As mentioned in Section~\ref{sec:background}, we have three types of observation (trap findings, berry infestation and egg findings) which all indicate whether D. suzukii occurs in a specific area. All of these observations serve as indicators that an area is infested, thus allowing us to combine them into a single measurement by first normalizing them to range $[0, 1]$ and afterwards summing them up into a single feature, subsequently referred to as \textit{observations}. To cope with irregular samplings of measurements, we averaged the number of observations per station per month. The resulting distribution is right-skewed, with most values being 0, meaning that for most stations we observe no infestation by D. suzukii in a month. To still be able to differentiate between severely infested stations and weakly or non-infested stations, we decided to set the 80~\% percentile as an experimental threshold to classify our stations. This threshold may be changed later requiring a retraining of our model, but otherwise not affecting the later steps of the classification and the usage of our application. In total we have a training set consisting of 3860 instances. Using the 80~\% percentile of the average \textit{observations} per month to partition our data into weakly (negative) and severely infested (positive) areas gives us a data set with 735 positive and 3125 negative instances. 


We enriched these instances, by adding information about the environmental surroundings of each station. First, we added the height information, which we extracted from ASTER. Second, we added the surrounding land usage information. Since a local spread is possible by D. suzukii itself, we extracted the land usage information in a 5~km radius around each station. Finally, we have an 85 dimensional feature vector for each instance, consisting of the month of the year, the station height, and the surrounding land usage.

Using this partitioning we end up with a rather imbalanced data set with four times as many negative examples as positive ones. This can cause problems since many machine learning algorithms depend on the assumption that the given data set is balanced~\cite{liu2010robust}. Although machine learning techniques exist which can deal with imbalanced data sets, such as the \textit{Robust Decision Trees} of Liu et al.~\cite{liu2010robust}, we want to employ ensemble-based classification, which is a combination of different classifiers. This allows us to improve the classification performance~\cite{rokach2010ensemble} and also to model the uncertainty of our classification, which aids people in making more informed decisions~\cite{skeels2010revealing}. This requires the creation of a balanced data set, which we can achieve by either using undersampling of the majority class or oversampling of the minority class. Undersampling can be achieved by stratified sampling using the infestation class as strata. However, this would remove instances from our already small data set. To avoid this, we employ oversampling of the minority class using the Synthetic Minority Over-sampling Technique (SMOTE)~\cite{chawla2002smote}. SMOTE picks pairs of nearest neighbors in the minority class and creates artificial instances by randomly placing a point on the line between the nearest neighbors until the data is balanced. Thus, allowing us to employ default machine learning algorithms. 

\subsection{Ensemble-based Classifier Training}

For training the classifiers we use the state-of-the-art data mining systems KNIME~\cite{BCDG+07} and WEKA~\cite{hall2009weka}. We use a selection of well-known machine learning techniques such as Decision Tree, Random Forest, Multilayer Perceptron, Probabilistic Neural Network from KNIME and k-nearest neighbor classifiers ($k\in[1, 5]$), LibSVM~\cite{CC01a}, Bayesian network, locally weighted learning and K* from WEKA. This selection was determined in an experimental evaluation of available classifiers in KNIME and WEKA, and might be extended later. In order to support our decision to employ ensemble-based classification to improve the classification performance, we first need a baseline measurement. We performed a 10-fold cross validation of each of the classifiers mentioned in the previous paragraph and found that the 1-nearest neighbor classifier achieved the best performance, with a mean Cohen's $\kappa$ score of 0.919. The other classifiers achieved Cohen's $\kappa$ scores between 0.50 and 0.89, as shown in Table~\ref{tbl:evaluation}, which are according to Altman~\cite{altman1990practical} moderate to very good agreement between the prediction and actual class. To test if ensemble-based classification could achieve better results, we used stacking~\cite{wolpert1992stacked}. Here a logistic regression model is trained which uses the prediction of all previously trained classifiers as inputs to make the final prediction, as suggested by Ho et al.~\cite{ho1994decision}. Using this approach we achieved a Cohen's $\kappa$ score of 0.927, which is only marginally better than the 1-nearest neighbor classifier. Nevertheless, we are now able to model the uncertainty of our prediction, which according to Skeels~\cite{skeels2010revealing} is important for decision-making.

\begin{table}[!ht]
	\centering
	\begin{tabular}{@{}r|l@{}}
        \toprule
        \textbf{Classifier} & \textbf{Cohen's $\kappa$} \\ \midrule
          \underline{Ensemble-based Classification}   &   \underline{0.927}\\
          1-NN Classifier	&	0.919\\
          K*	&	0.887\\
          Random Forest	&	0.866\\
          Decision Tree	&	0.864\\
          2-NN Classifier	&	0.812\\
          3-NN Classifier	&	0.758\\
          4-NN Classifier	&	0.728\\
          Probabilistic Neural Network	&	0.718\\
          5-NN Classifier	&	0.693\\
          LibSVM	&	0.593\\
          Multilayer Perceptron	&	0.545\\
          Bayesian network	&	0.512\\
          Locally weighted learning	&	0.504\\
	\end{tabular}
    \caption{The ensemble-based classification achieved the best results, in accordance with the study by Rokach~\cite{rokach2010ensemble}}
    \label{tbl:evaluation}
\end{table}


\section{Drosophigator: Visual Analysis of Spatio-Temporal Event Predictions}


\mbox{}\\
\vspace*{-0.3cm}
\begin{figure}[!htb]
	\centering
	\includegraphics[width=0.9\columnwidth]{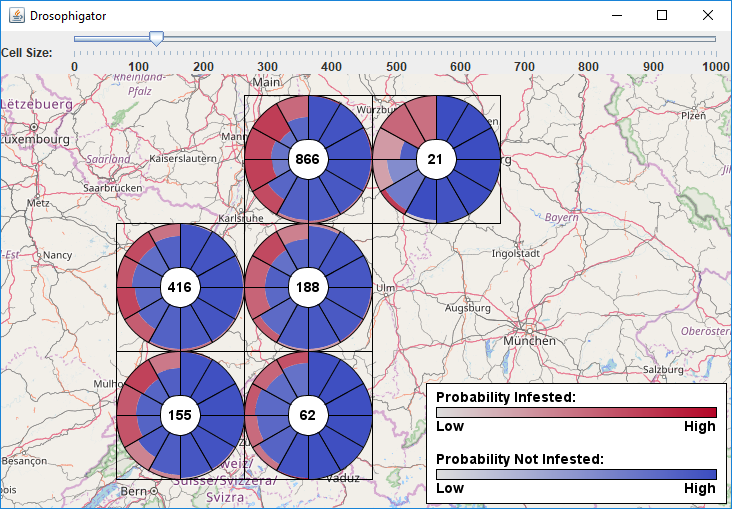}
	\caption{An overview of the \textit{Drosophigator} application for the visual analysis of spatio-temporal event predictions.}
	\label{fig:drosophigator}
\end{figure}

Just providing the users with the raw results of our prediction is not sufficient - on the one hand because we have over 20.000 predictions for all months and vineyards in Baden-W\"{u}rttemberg and on the other hand because the spatial context is not interpretable which makes it hard for experts to integrate their domain knowledge into the analysis process. Thus, we need visualization to help experts to easily identify spatial and temporal patterns, select areas of interest and get detailed information on the uncertainty of our model. To achieve this, we follow the visual information seeking mantra of Ben Shneiderman: ``Overview first, zoom and filter, details on demand''~\cite{shneiderman1996eyes}.

\begin{figure*}
	\centering
	\subfloat[Each time-segment can be divided into multiple visualization areas, which correspond to the number of possible events. The more interesting events to observe are placed on the outside, to be more easily visible. The ratio indicator is used to illustrate which event is most likely to occur.]
	{\includegraphics[height=0.21\linewidth,page=2]{pictures/glyph_visualization.pdf}}\hspace{2mm}
	\subfloat[Time-segments are ordered clockwise. They can represent different time units, from hours, days, months or years. Each time-segment can be used to visualize the event prediction for the specified time frame.]
	{\includegraphics[height=0.21\linewidth]{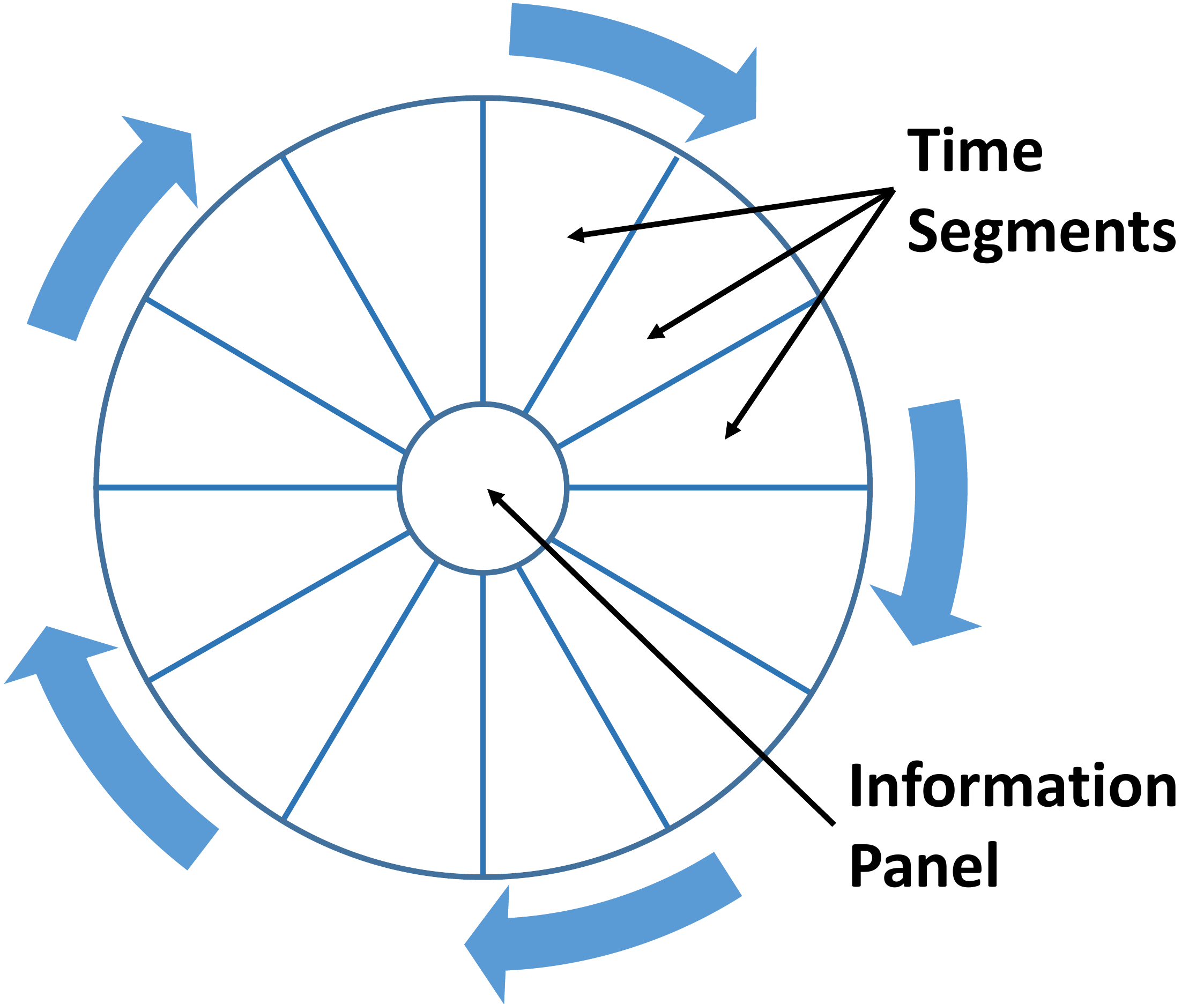}}\hspace{2mm}
    \subfloat[Example Glyph Visualization showing the ratio of infested and not infested vineyards, as well the uncertainty of the prediction for one year.]
	{\includegraphics[height=0.21\linewidth]{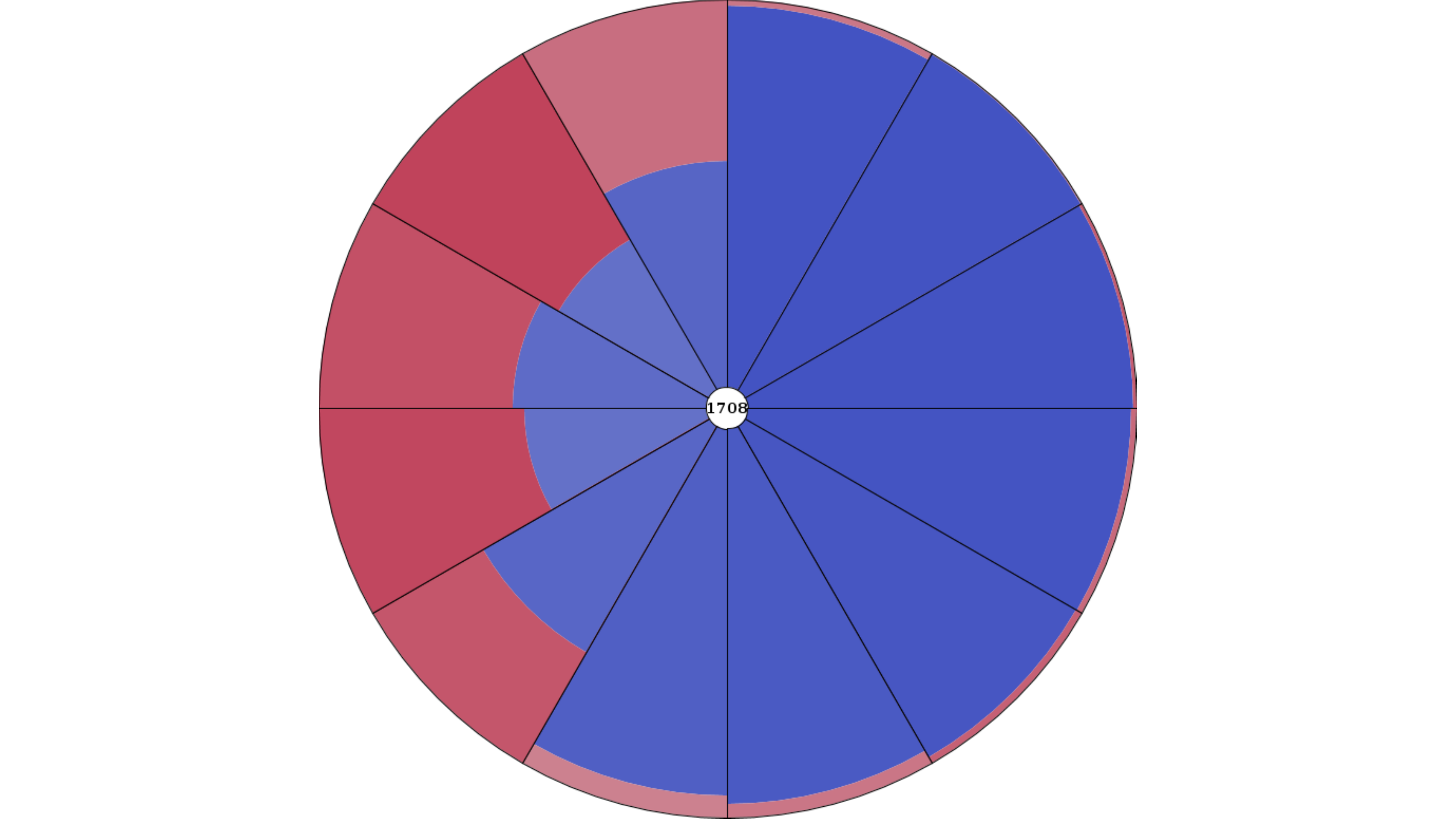}}
	\caption{Sketch of a single time-segment (a), the resulting glyh-based temporal event prediction visualization (b) and a real glyph example (c).}
	\label{img:glyph}
\end{figure*}

As depicted in Figure~\ref{img:vineyards_bawu}, the distribution of vineyards in Baden-W\"{u}rttemberg is very sparse and several clusters can be immediately spotted. According to the goals of the analysis system stated in Section~\ref{sec:background}, we need to offer the user an overview first before interactively digging deeper into several spatial areas according to the domain knowledge of the user who could be either a farmer, vintner or researcher. Existing related systems such as BirdVis~\cite{ferreira2011birdvis} offer heat map overview visualizations. However, as we want to investigate the distribution of a species over time, we designed a map overlay consisting of several glyphs. This partially preserves the geographic context while the position of a glyph can be used to encode additional information to a certain degree. 

The glyph proposed in this paper has a fixed size such that its visual encodings can be interpreted well. Furthermore, this allows the user to compare various glyphs. The position and size of each glyph is determined by a regular grid drawn onto the map. Each grid cell aggregates all vineyards within the respective grid cell. Since the vineyards' polygons might appear very small at the most zoom levels of the map, we decide to use the center of a vineyard's polygon to assign it to its respective grid cell. Moreover we want to enable the user to drill-down. This is realized by a semantic zoom into a specific spatial area which enhances well known zooming techniques offered by the most interactive maps. The semantic of the zoom lies within the position and size of the glyphs. On a zoom the size of a glyph stays the same while the map is enlarged or reduced which consequently results in a split of each grid cell such that one glyph covers a smaller spatial area and thus represents a more detailed view. Under certain circumstances, it also may not be enough to be limited to the predefined size of a glyph. Therefore the system offers the possibility to adjust the size of a glyph within a reasonable range by a common slider which is placed above the map view. This allows a seamless investigation of differently sized areas without the necessity to change the map's view-port by zooming in or out a specific area.


Extensive work on glyphs has been done in the past which we used as guidelines in order to design the final glyph proposed in this paper. This includes e.g. the work by Fuchs et al.~\cite{Fuchs:2013glyph} and Borgo et al.~\cite{borgo2013glyph}.
We are dealing with time series data, specifically with periodic time series, namely months.
Additionally, a user might want to test time related hypotheses. This would support a radial layout of the glyph which facilitates a seamless comparison of neighbored time periods since they are visually neighbored as well. Therefore, we decide to arrange each month within a circle which results in a circle consisting of twelve circle segments (time segments) such that we utilize the visual metaphor of a clock. We therefore chose January to be represented by the first time segment (from 12 to 1 o'clock) and then increase monthly clockwise. Such a clock metaphor was also used by Fischer et al. to visualize periodic time data~\cite{FFM12a}. Moreover, we use the center of the circle to represent additional time-independent information, namely the number of vineyards the respective glyph represents. The basic design of the implemented glyph is depicted in Figure \ref{img:glyph}~(b).

\begin{figure}[!ht]
	\centering
	\includegraphics[width=\columnwidth,page=1]{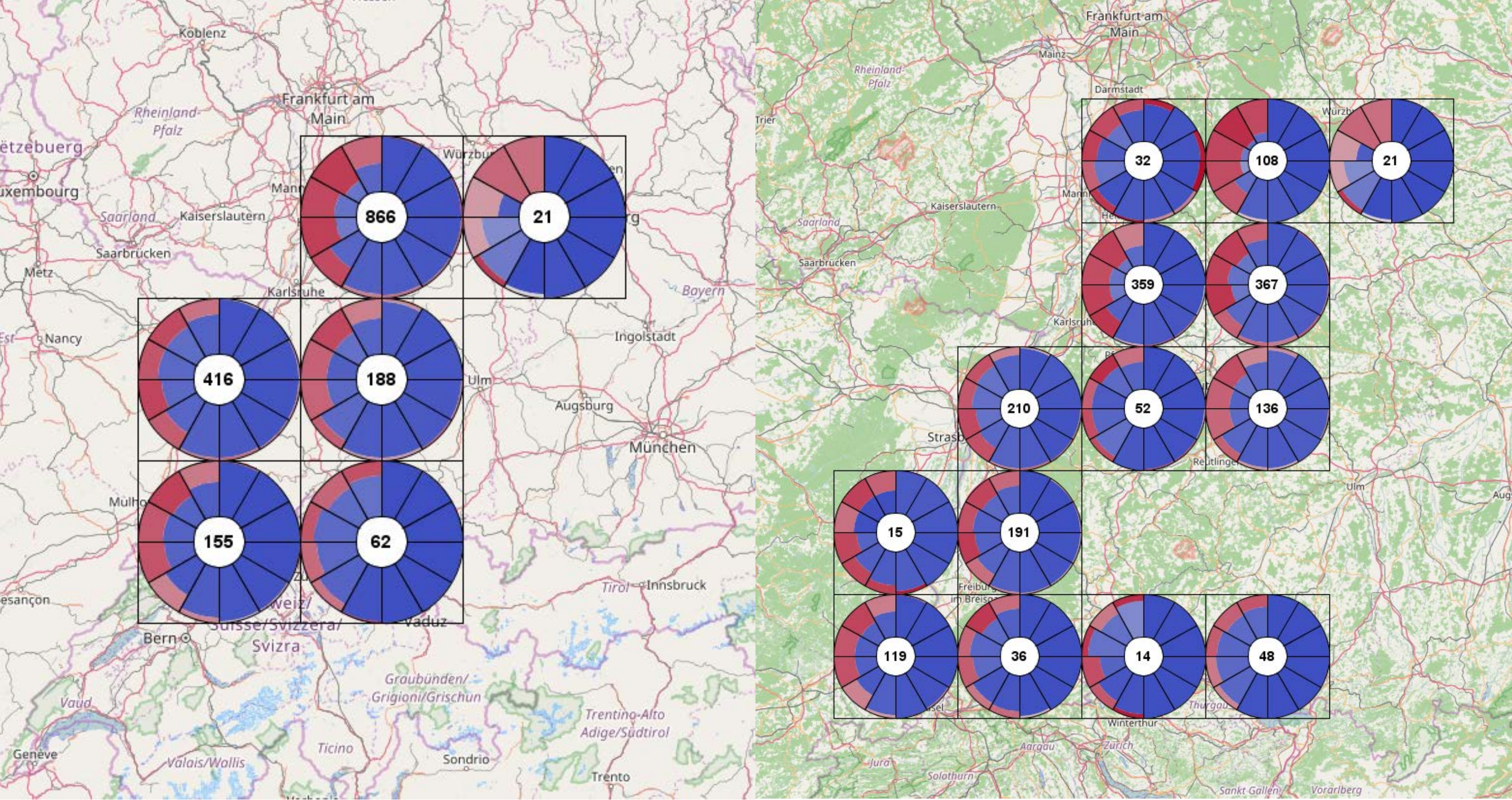}
    \caption{Overview of our glyph-based visualization. For each cell the predictions and their uncertainties are averaged per time-segment and visualized in our glyph. We provide zoom-and-filter capabilities by allowing the user to zoom in and out of the map, as well as manually adjusting the cell size to either get a broader or more detailed view on the underlying data.}
    \label{img:tool_overview}
\end{figure}

Another goal is to visualize whether a certain region is endangered or not. Consequently, we visually encode the classification results of a specific month represented by its time segment. The basic design of a time segment is depicted in Figure~\ref{img:glyph}~(a). Therefore, we make use of the interior of the respective time segment to represent the classification results of the ensemble-classifiers applied in Section~\ref{sec:classification}. For each month we have a distribution of safe and endangered vineyards (according to the classification). Since the number of vineyards stay the same over all months for each glyph, we fill the area of the time segments according to the ratio of the binary outcome. This technique results in a radial glyph similar to a stacked bar chart showing fractions of the whole. 
To be able to distinguish the outcome we use the colors red (endangered) and blue (not endangered) as derived from the warm-cold color scale~\cite{moreland2009diverging}. Additionally, a probability is assigned to each outcome (endangered or not) and is of high relevance since it represents the (un-)certainty and thus helps to find out where to place additional measurement stations. The average probability of a given outcome is encoded using the respective half of the warm-cold color scale, such that a high probability/certainty results in a stronger color tone while a low probability on the other hand is represented by a very weak color tone. A very strong red color exemplary means that there is a very high probability of endangered vineyards within the respective month (time segment) and area (glyph location). This enables the user to immediately spot potentially new measurement areas.  An overview of the realized glyph representing all vineyards in Baden-W\"{u}rttemberg is presented in Figure~\ref{img:tool_overview}. It can immediately be observed that there is a general trend as the number of endangered vineyards (red) is increasing until late summer and then decreasing again. This observation corroborates the hypothesis that the D. suzukii may only survive in a relative stable environment regarding temperature such that it dries out in the summer and freezes in winter months.

\begin{figure}[!ht]
	\centering
	\includegraphics[width=\columnwidth,page=2]{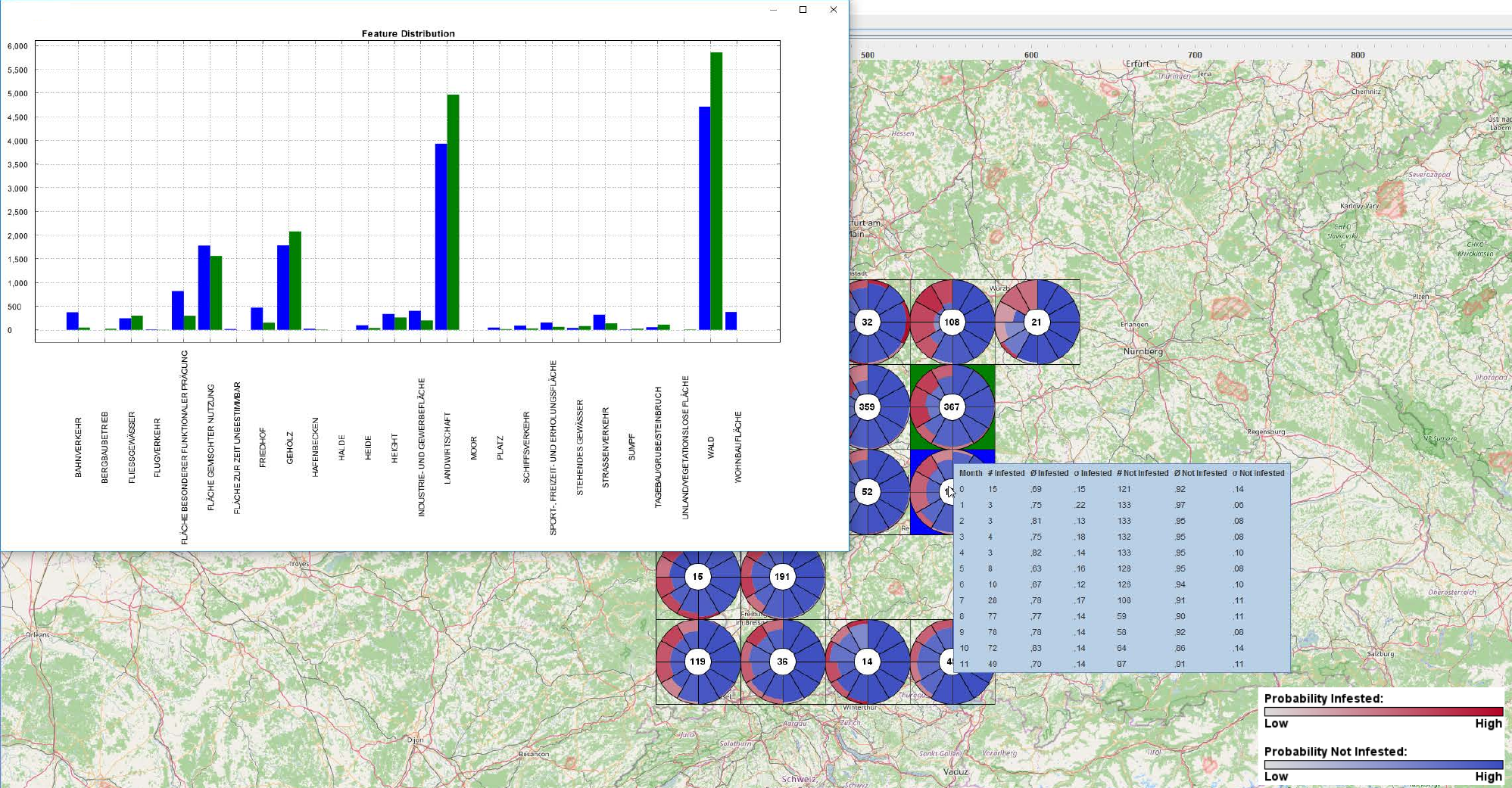}
    \caption{Overview of our detail-on-demand visualizations. We provide tooltips, detailing the number of infested and not-infested areas per month per cell, as well as statistics about the measured uncertainty, such as average and standard deviation. Additionally, we provide bar chart visualizations of the land usage. These can be used to compare the average land usage of the vineyards contained in different cells. The bar colors correspond to the highlight colors of the selected cells.}
    \label{img:tooltip}
\end{figure}

It is possible to zoom until there is only one glyph for each vineyard visible on the display. Nevertheless, for further analysis regarding the question ``why'' the ensemble-classification reports a specific result, it is necessary to show the input feature vectors of the classification in an easily understandable way. Moreover, the ability to compare areas/glyphs with each other regarding the feature vectors enables the visual detection of several environment compositions (e.g., surrounding land use and height) that contribute crucial to a given classification result. Furthermore, the visual representation of the feature vectors builds additional trust in the classification result as a user may want to trace why a certain area has a specific result. Therefore, we enable the user to select one or more glyphs to show additional details on demand. These details of the input features are shown within a bar chart as depicted in Figure~\ref{img:tooltip}. The color of a bar is directly associated with the dedicated glyph which is highlighted in the same color to visually link them.

\section{Use Cases}

In this section, we want to highlight how visualization can help domain experts to gain insights about the spread dynamics of D. suzukii. We show the usefulness of our system by demonstrating how domain experts can investigate hypotheses using Drosophigator. Therefore, we investigated two recently proposed assumptions about the time of infestation~\cite{jki} and the influences of environmental factors~\cite{Pelton2016}.  

\begin{figure}[!ht]
	\centering
	\includegraphics[width=\columnwidth]{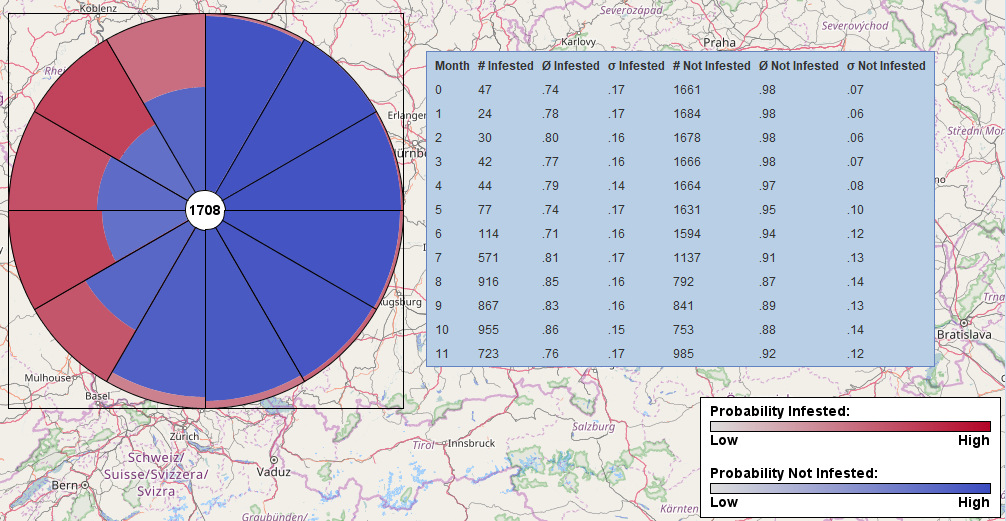}
    \caption{Overview glyph-visualization of all vineyards in Baden-W\"{u}rttemberg. The development over the time-segments shows that the severity of infestation and the certainty of our prediction increases in late summer and stays high until the end of the year. This corroborates the hypothesis of the JKI~\cite{jki}.}
    \label{fig:more_kef_in_late_summer}
\end{figure}

The JKI states as a general rule, that the number of findings increases with decreasing temperatures in late summer and stays high until November or later if there are no cold snaps~\cite{jki}. To investigate this hypothesis we create an overview of all available predictions. We increase the cell size of our grid such that a single cell covers all vineyards in Baden-W\"{u}rttemberg. The resulting glyph-visualization is shown in Figure~\ref{fig:more_kef_in_late_summer}. In the visualization, as well as in the detailed tooltip, we can see that the number of infestation is marginal in the first half of the year. However, there is a strong increase in the predicted number of infestation and diminishing uncertainty starting in August until December. This observation is consistent with the hypothesis of the JKI.   

\begin{figure}[!ht]
	\centering
	\includegraphics[width=\columnwidth]{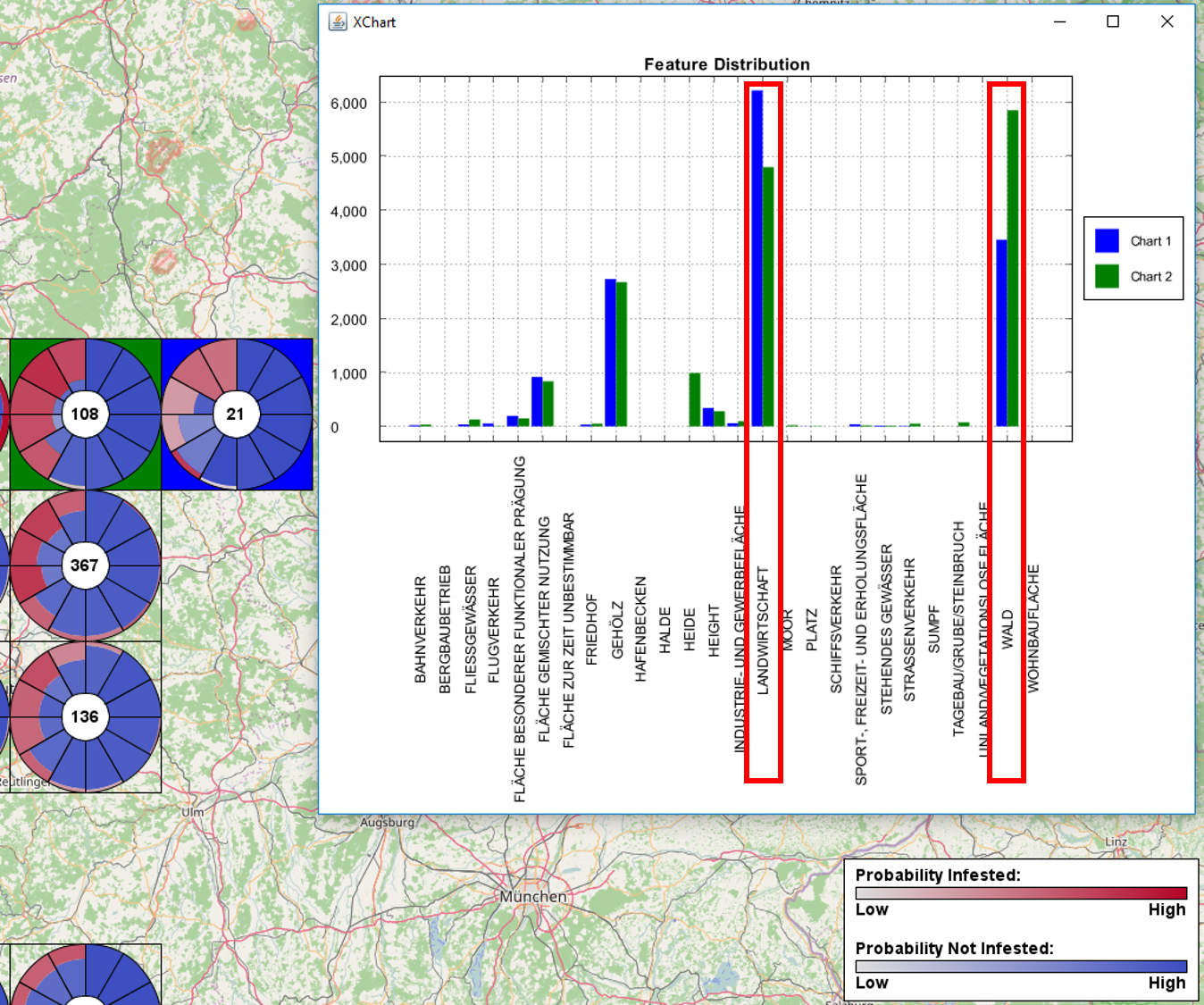}
    \caption{Comparison of the vineyards contained in two neighboring cells. The left cell (green) exhibits an earlier infestation by D. suzukii that the right cell (blue). The detail visualization shows, that the vineyards in the left cell have more surrounding woodland (\textit{Wald}) than those in the right cell. This finding strongly supports the hypothesis of Pelton et al.~\cite{Pelton2016}.}
    \label{fig:earlier_kef_activation_due_to_wood}
\end{figure}

A recent two-year field study of Pelton et al.~\cite{Pelton2016} suggests, that high amounts of surrounding woodland are correlated with earlier an infestation of D. suzukii. Using a finer grid resolution, we compare two neighboring grid cells, as shown in Figure~\ref{fig:earlier_kef_activation_due_to_wood}. The left cell (highlighted in green) shows an earlier infestation than the right cell (highlighted in blue). We compare the land usage of vineyards contained in both cells using our detail-on-demand bar chart visualization. We can identify that the left cell, which exhibits an earlier infestation, has a larger amount of woodlands in the surrounding area, while the right has more agricultural areas in its surrounding. These observations support the hypothesis of Pelton et al.

In these two use-cases, we have demonstrated the capabilities of our tool. Following the information-seeking mantra of Shneidermann using our glyph-based visualization of the ensemble-based predictions, as well as the uncertainty of the prediction, allows us to make observations supporting hypotheses of researchers about the spread dynamics of D. suzukii.

\section{Discussion and Future Work}

The approach presented in this paper is currently very application-driven. However, first feedback from experts is very promising. There are still limitations, which need to be addressed in future work. For instance, the inability to specify the time-granularity makes it currently impossible to investigate the spread dynamics over the course of multiple years. This may require an extension of our visualization, for example, by following a similar approach as proposed by Carlis and Konstan~\cite{carlis1998interactive}. Further potential improvements of our visualization include switching from a regular grid to a dynamic aggregation approach, avoiding overemphasis of large areas in the glyph with low confidence following the guidelines of Ferreira et al.~\cite{ferreira2014sample}. Additionally, we plan to reduce the occlusion introduced through our glyph by investigating advanced alternative visualization techniques. Eventually, we need to incorporate other data sources, especially meteorological data. This would allow the investigation of the effects of late frost or severe heat and drought on the spread of D. suzukii, since adult flies are very susceptible to weather~\cite{jakobs2015adult}.

For additional future work we aim to investigate the applicability of our system for spatio-temporal event analysis of other species. One particular use case will be the Global Initiative for Honey Bee Health (GIHH) launched by the CSIRO in 2015 \cite{gihh}, which aims to collect scientific evidence of honey bee population decline through global collaboration. Towards this end, microsensors are attached to the bees to record their activity from which predictors of health are inferred. A visual analytics framework \cite{engelke2016vizzzbees} is being developed that facilitates interactive analysis of the microsensing data and aids in finding correlates with environmental factors that may impact on bee health. The system we are presenting here is considered a valuable means of visually interrogating the health predictors and their related uncertainties on a global scale. 

\section{Conclusion}

In this paper we presented our application \textit{Drosophigator} which enables researchers in the field of viticulture and biology to investigate the spread dynamics of invasive species. Using data provided by the WBI we trained an ensemble of classifiers to identify places and times which are susceptible to infestation by D. suzukii. Using our glyph-based visualization we allow a visual analysis of these spatio-temporal event predictions. We demonstrated the capabilities of our approach in two use-cases, where we show how our tool can be used to investigate hypothesis about the spread of D. suzukii.

\acknowledgments{
\label{sec:ack} The project BigGIS (reference number: 01IS14012) is funded by the Federal
Ministry of Education and Research (BMBF) within the frame of the research
programme ``Management and Analysis of Big Data'' in ``ICT 2020 --
Research for Innovations''.}

\bibliographystyle{abbrv-doi}

\bibliography{template}

\begin{thebibliography}{10}

\bibitem{aster}
{ASTER Global Digital Elevation Map}.
\newblock \url{https://asterweb.jpl.nasa.gov/gdem.asp}.
\newblock Accessed: 2017-07-06.

\bibitem{jki}
{Biologie der Drosophila Suzukii}.
\newblock \url{http://drosophila.jki.bund.de/index.php?menuid=28}.
\newblock Accessed: 2017-07-06.

\bibitem{gihh}
{Global Initiative for Honey Bee Health (GIHH)}.
\newblock https://research.csiro.au/gihh/about/.
\newblock Accessed: 07/03/2017.

\bibitem{altman1990practical}
D.~G. Altman.
\newblock {\em Practical statistics for medical research}.
\newblock CRC press, 1990.

\bibitem{asplen2015invasion}
M.~K. Asplen, G.~Anfora, A.~Biondi, V.~M. Walton, et~al.
\newblock Invasion biology of spotted wing drosophila (drosophila suzukii): a
  global perspective and future priorities.
\newblock 2015.

\bibitem{BCDG+07}
M.~R. Berthold, N.~Cebron, F.~Dill, T.~R. Gabriel, T.~K\"{o}tter, T.~Meinl,
  P.~Ohl, C.~Sieb, K.~Thiel, and B.~Wiswedel.
\newblock {KNIME}: The {K}onstanz {I}nformation {M}iner.
\newblock In {\em Studies in Classification, Data Analysis, and Knowledge
  Organization (GfKL 2007)}. Springer, 2007.

\bibitem{borgo2013glyph}
R.~Borgo, J.~Kehrer, D.~H. Chung, E.~Maguire, R.~S. Laramee, H.~Hauser,
  M.~Ward, and M.~Chen.
\newblock Glyph-based visualization: Foundations, design guidelines, techniques
  and applications.
\newblock In {\em Eurographics (STARs)}, pp. 39--63, 2013.

\bibitem{carlis1998interactive}
J.~V. Carlis and J.~A. Konstan.
\newblock Interactive visualization of serial periodic data.
\newblock In {\em Proceedings of the 11th annual ACM symposium on User
  interface software and technology}, pp. 29--38. ACM, 1998.

\bibitem{CC01a}
C.-C. Chang and C.-J. Lin.
\newblock {LIBSVM}: A library for support vector machines.
\newblock {\em ACM Transactions on Intelligent Systems and Technology},
  2:27:1--27:27, 2011.
\newblock Software available at \url{http://www.csie.ntu.edu.tw/~cjlin/libsvm}.

\bibitem{chawla2002smote}
N.~V. Chawla, K.~W. Bowyer, L.~O. Hall, and W.~P. Kegelmeyer.
\newblock Smote: synthetic minority over-sampling technique.
\newblock {\em Journal of artificial intelligence research}, 16:321--357, 2002.

\bibitem{cini2014tracking}
A.~Cini, G.~Anfora, L.~Escudero-Colomar, A.~Grassi, U.~Santosuosso, G.~Seljak,
  and A.~Papini.
\newblock Tracking the invasion of the alien fruit pest drosophila suzukii in
  europe.
\newblock {\em Journal of Pest Science}, 87(4), 2014.

\bibitem{cini2012review}
A.~Cini, C.~Ioriatti, G.~Anfora, et~al.
\newblock A review of the invasion of drosophila suzukii in europe and a draft
  research agenda for integrated pest management.
\newblock {\em Bulletin of insectology}, 65(1):149--160, 2012.

\bibitem{cormier2015exclusion}
D.~Cormier, J.~Veilleux, and A.~Firlej.
\newblock Exclusion net to control spotted wing drosophila in blueberry fields.
\newblock {\em IOBC-WPRS Bull}, 109:181--184, 2015.

\bibitem{daane2016first}
K.~M. Daane, X.-G. Wang, A.~Biondi, B.~Miller, J.~C. Miller, H.~Riedl, P.~W.
  Shearer, E.~Guerrieri, M.~Giorgini, M.~Buffington, et~al.
\newblock First exploration of parasitoids of drosophila suzukii.
\newblock {\em Journal of pest science}, 89(3):823--835, 2016.

\bibitem{del2017cost}
E.~Del~Fava, C.~Ioriatti, and A.~Melegaro.
\newblock Cost-benefit analysis of controlling the spotted wing drosophila
  (drosophila suzukii (matsumura)) spread and infestation of soft fruits in
  trentino, northern italy.
\newblock {\em Pest Management Science}, 2017.

\bibitem{engelke2016vizzzbees}
U.~Engelke, P.~Marendy, F.~Susanto, R.~Williams, S.~Mahbub, H.~Nguyen, and
  P.~de~Souza.
\newblock A visual analytics framework to study honey bee behaviour.
\newblock In {\em Proceedings of the 2nd IEEE Data Science and Systems
  Conference (DSS)}, pp. 1504--1511. IEEE, 2016.

\bibitem{ferreira2014sample}
N.~Ferreira, D.~Fisher, and A.~C. Konig.
\newblock Sample-oriented task-driven visualizations: allowing users to make
  better, more confident decisions.
\newblock In {\em Proceedings of the SIGCHI Conference on Human Factors in
  Computing Systems}, pp. 571--580. ACM, 2014.

\bibitem{ferreira2011birdvis}
N.~Ferreira, L.~Lins, D.~Fink, S.~Kelling, C.~Wood, J.~Freire, and C.~Silva.
\newblock Birdvis: Visualizing and understanding bird populations.
\newblock {\em IEEE Transactions on Visualization and Computer Graphics},
  17(12):2374--2383, 2011.

\bibitem{FFM12a}
F.~Fischer, J.~Fuchs, and F.~Mansmann.
\newblock {ClockMap: Enhancing Circular Treemaps with Temporal Glyphs for
  Time-Series Data}.
\newblock In M.~Meyer and T.~Weinkauf, eds., {\em Proceedings of the
  Eurographics Conference on Visualization (EuroVis 2012 Short Papers)}, pp.
  97--101. Vienna, Austria, 2012. doi: {{%
10\hspace{.1pt}\discretionary{.}{%
}{.}\hspace{.4pt}2312\discretionary{/}{%
}{/}PE\discretionary{/}{%
}{/}EuroVisShort\discretionary{/}{%
}{/}EuroVisShort2012\discretionary{/}{%
}{/}097\discretionary{%
}{-}{-}101}}


\bibitem{Fuchs:2013glyph}
J.~Fuchs, F.~Fischer, F.~Mansmann, E.~Bertini, and P.~Isenberg.
\newblock Evaluation of alternative glyph designs for time series data in a
  small multiple setting.
\newblock In {\em Proceedings of the SIGCHI Conference on Human Factors in
  Computing Systems}, CHI '13, pp. 3237--3246. ACM, New York, NY, USA, 2013.

\bibitem{grunreich1992atkis}
D.~Gr{\"u}nreich.
\newblock Atkis-a topographic information system as a basis for gis and digital
  cartography in germany.
\newblock {\em From Digital Map Series to Geo-Information Systems, Geologisches
  Jarhrbuch Series A. Hannover, Germany: Federal Institute of Geosciences and
  Resources}, 1992.

\bibitem{hall2009weka}
M.~Hall, E.~Frank, G.~Holmes, B.~Pfahringer, P.~Reutemann, and I.~H. Witten.
\newblock The weka data mining software: an update.
\newblock {\em ACM SIGKDD explorations newsletter}, 11(1):10--18, 2009.

\bibitem{haye2016current}
T.~Haye, P.~Girod, A.~Cuthbertson, X.~Wang, K.~Daane, K.~Hoelmer, C.~Baroffio,
  J.~Zhang, and N.~Desneux.
\newblock Current swd ipm tactics and their practical implementation in fruit
  crops across different regions around the world.
\newblock {\em Journal of pest science}, 89(3):643--651, 2016.

\bibitem{ho1994decision}
T.~K. Ho, J.~J. Hull, and S.~N. Srihari.
\newblock Decision combination in multiple classifier systems.
\newblock {\em IEEE transactions on pattern analysis and machine intelligence},
  16(1):66--75, 1994.

\bibitem{jakobs2015adult}
R.~Jakobs, T.~D. Gariepy, and B.~J. Sinclair.
\newblock Adult plasticity of cold tolerance in a continental-temperate
  population of drosophila suzukii.
\newblock {\em Journal of insect physiology}, 79:1--9, 2015.

\bibitem{kaiser2015drosphila}
L.~Kaiser.
\newblock Drosophila suzukii - wissenswertes und l\"{o}sungsans\"{a}tze aus der
  forschung.
\newblock 2015.

\bibitem{liu2010robust}
W.~Liu, S.~Chawla, D.~A. Cieslak, and N.~V. Chawla.
\newblock A robust decision tree algorithm for imbalanced data sets.
\newblock In {\em Proceedings of the 2010 SIAM International Conference on Data
  Mining}, pp. 766--777. SIAM, 2010.

\bibitem{moreland2009diverging}
K.~Moreland.
\newblock Diverging color maps for scientific visualization.
\newblock {\em Advances in Visual Computing}, pp. 92--103, 2009.

\bibitem{Pelton2016}
E.~Pelton, C.~Gratton, R.~Isaacs, S.~Van~Timmeren, A.~Blanton, and
  C.~Gu{\'e}dot.
\newblock Earlier activity of drosophila suzukii in high woodland landscapes
  but relative abundance is unaffected.
\newblock {\em Journal of Pest Science}, 89(3):725--733, Jul 2016. doi: {{%
10\hspace{.1pt}\discretionary{.}{%
}{.}\hspace{.4pt}1007\discretionary{/}{%
}{/}s10340\discretionary{%
}{-}{-}016\discretionary{%
}{-}{-}0733\discretionary{%
}{-}{-}z}}


\bibitem{rokach2010ensemble}
L.~Rokach.
\newblock Ensemble-based classifiers.
\newblock {\em Artificial Intelligence Review}, 33(1):1--39, 2010.

\bibitem{rossi2016multiple}
M.~V. Rossi-Stacconi, R.~Kaur, V.~Mazzoni, L.~Ometto, A.~Grassi,
  A.~Gottardello, O.~Rota-Stabelli, and G.~Anfora.
\newblock Multiple lines of evidence for reproductive winter diapause in the
  invasive pest drosophila suzukii.
\newblock {\em Journal of pest science}, 89(3):689--700, 2016.

\bibitem{shneiderman1996eyes}
B.~Shneiderman.
\newblock The eyes have it: A task by data type taxonomy for information
  visualizations.
\newblock In {\em Visual Languages, 1996. Proceedings., IEEE Symposium on}, pp.
  336--343. IEEE, 1996.

\bibitem{skeels2010revealing}
M.~Skeels, B.~Lee, G.~Smith, and G.~G. Robertson.
\newblock Revealing uncertainty for information visualization.
\newblock {\em Information Visualization}, 9(1):70--81, 2010.

\bibitem{wang2016foraging}
X.-G. Wang, G.~Ka{\c{c}}ar, A.~Biondi, and K.~M. Daane.
\newblock Foraging efficiency and outcomes of interactions of two pupal
  parasitoids attacking the invasive spotted wing drosophila.
\newblock {\em Biological Control}, 96:64--71, 2016.

\bibitem{wiman2014integrating}
N.~G. Wiman, V.~M. Walton, D.~T. Dalton, G.~Anfora, H.~J. Burrack, J.~C. Chiu,
  K.~M. Daane, A.~Grassi, B.~Miller, S.~Tochen, et~al.
\newblock Integrating temperature-dependent life table data into a matrix
  projection model for drosophila suzukii population estimation.
\newblock {\em PLoS One}, 9(9):e106909, 2014.

\bibitem{wolpert1992stacked}
D.~H. Wolpert.
\newblock Stacked generalization.
\newblock {\em Neural networks}, 5(2):241--259, 1992.

\end{thebibliography}
\end{document}